\documentclass[conference]{IEEEtran}
\IEEEoverridecommandlockouts

\usepackage{cite}
\usepackage{amsmath,amssymb,amsfonts}
\usepackage{algorithmic}
\usepackage{graphicx}
\usepackage{textcomp}
\usepackage{xcolor}
\usepackage{float}

\usepackage{tikz}
\usepackage{pgfplots}
\pgfplotsset{compat=1.18}
\usepackage{pgfplotstable}
\def\BibTeX{{\rm B\kern-.05em{\sc i\kern-.025em b}\kern-.08em
    T\kern-.1667em\lower.7ex\hbox{E}\kern-.125emX}}

\usepackage{pifont}
\usepackage{subfig}
\usepackage{colortbl}
\usepackage{braket}
\usepackage{comment}
\usepackage{hyperref}
\usepackage{stfloats}

\newcommand{\CellWithForceBreak}[2][c]{
\begin{tabular}[#1]{@{}c@{}}#2\end{tabular}}

\newcommand{\acronym}{CLOAQ}

\usepackage[compact]{titlesec}
\titlespacing*{\section} {0pt}{1ex}{0ex} 
\titlespacing*{\subsection} {0pt}{0.5ex}{0ex}
\titlespacing*{\subsubsection} {0pt}{0.5ex}{0ex}

\usepackage{balance}
\usepackage{xcolor}
\usepackage[ruled,vlined,noend]{algorithm2e}

\SetCommentSty{mycommfont}
\SetAlFnt{\small}
\SetAlCapFnt{\small}
\SetAlCapNameFnt{\small}

\usepackage{soul}
\def\BibTeX{{\rm B\kern-.05em{\sc i\kern-.025em b}\kern-.08em
    T\kern-.1667em\lower.7ex\hbox{E}\kern-.125emX}}

\title{\acronym: Combined Logic and Angle Obfuscation for Quantum Circuits}
\author{
    \IEEEauthorblockN{Vincent Langford\IEEEauthorrefmark{1}, Shihan Zhao\IEEEauthorrefmark{1}, Hongyu Zhang\IEEEauthorrefmark{1}, Ben Dong\IEEEauthorrefmark{2}, Qian Wang\IEEEauthorrefmark{2}, Anees Rehman\IEEEauthorrefmark{3}, Yuntao Liu\IEEEauthorrefmark{1}}
    \IEEEauthorblockA{\IEEEauthorrefmark{1}Department of Electrical and Computer Engineering, Lehigh University, PA, USA}
    \IEEEauthorblockA{\IEEEauthorrefmark{2}Department of Electrical Engineering, University of California, Merced, CA, USA}
    \IEEEauthorblockA{\IEEEauthorrefmark{3}Institute of Sensors, Signals and Systems, Heriot-Watt University, Edinburgh, UK}
    \IEEEauthorblockA{\IEEEauthorrefmark{1}\{val26, shz425, hoz324, yule24\}@lehigh.edu, \IEEEauthorrefmark{2}\{cdong12, qianwang\}@ucmerced.edu, \IEEEauthorrefmark{3}ar4033@hw.ac.uk}

}

\date{}

\begin{document}

\maketitle



\begin{abstract}
In the realm of quantum computing, quantum circuits serve as essential depictions of quantum algorithms, which are then compiled into executable operations for quantum computations. Quantum compilers are responsible for converting these algorithmic quantum circuits into versions compatible with specific quantum hardware, thus connecting quantum software with hardware. Nevertheless, untrusted quantum compilers present notable threats. They have the potential to result in the theft of quantum circuit designs and jeopardize sensitive intellectual property (IP).
In this work, we propose \acronym, a quantum circuit obfuscation (QCO) approach that hides the logic and the phase angles of selected gates within the obfuscated quantum circuit.
To evaluate the effectiveness of \acronym, we sample the input state uniformly from the Hilbert space of all qubits, which is more accurate than prior work that use all-$\ket{0}$ inputs.
Our results show that \acronym\ benefits from the synergy between logic and phase protections. Compared with prior QCO approaches using only one perspective, the combined method is more resilient to attacks and causes greater functional disruption when the unlocking key is incorrect.

\end{abstract}

\maketitle

\section{Introduction}

The rise of quantum computing heralds a new era of computational prowess and techniques. Despite notable progress in quantum computing, creating quantum circuits remains a resource-heavy task. It usually requires multiple iterations to achieve optimal performance. Therefore, the design of quantum circuits is immensely valuable and is considered crucial intellectual property (IP) \cite{aboy2022mapping}, susceptible to risks akin to traditional IC designs. This paper concentrates on the IP threats that quantum circuits face during the compilation phase. In the compilation phase, the compiler translates the original quantum circuit into a form understandable and executable by the quantum computer. This process involves replacing all gates with those supported by quantum computing hardware and inserting swap gates to ensure that the physical qubits' connections can realize the multi-qubit operations. Additionally, several optimizations can be applied to enhance the compiled circuits' performance. Nonetheless, this process is vulnerable, as malicious compilers may exploit and misuse the quantum designs, such as by implanting Trojans \cite{das2023trojannet, roy2024hardware} and forging quantum designs \cite{yang2024multi}. Given that quantum circuits constitute valuable IP \cite{aboy2022mapping}, it is crucial to safeguard them against unauthorized exploitation and IP infringement.

Multiple quantum circuit obfuscation (QCO) techniques have been proposed in prior work, including key-based techniques \cite{topaloglu2023quantum, liu2025eloqenhancedlockingquantum, rehman2025opaque} and split compilation \cite{saki2021split, wang2025tetrislock}.
Our work primarily focuses on key-based techniques, which use key bits to determine logic flips \cite{saki2021split, liu2025eloqenhancedlockingquantum} or phase angles \cite{rehman2025opaque}. Logic flips generally produce higher corruption, whereas phase-based obfuscation can accommodate more key bits per gate because of the continuous nature of phase angles, exponentially increasing the key search space.
In this paper, we explore the methodology to combine the benefits of the two.
Our major contributions are as follows:
\begin{itemize}
   \item We propose \acronym, a novel key-based QCO approach that controls both logic and phase using key values.
   \item We introduce an input state sampling approach to evaluate the overall functional impact of obfuscation on quantum circuit functionality, obtaining more representative results than current evaluation approaches that use all-$\ket{0}$ states.
   \item Our experiment results show that we can effectively lock the circuit and accurately recovered with minimal accuracy loss.  
\end{itemize}

\section{Background \& Related Work}

Several strategies have been investigated by researchers to safeguard quantum circuits from threats originating from untrusted compilers. These strategies include the integration of random reversible circuits \cite{das2023randomized, suresh2021short, das2024secure}, locking the circuits by adding extra key qubits \cite{topaloglu2023quantum, liu2025eloqenhancedlockingquantum}, encoding phase gate rotation angles in key bits \cite{rehman2025opaque}, and dividing the quantum circuit to compile each segment using separate compilers \cite{saki2021split, wang2025tetrislock}. In studies \cite{das2023randomized} and \cite{das2024secure}, a randomly generated reversible quantum circuit was embedded into the initial circuit, and post-compilation, the inverse of this random circuit was applied in the same location to revert functionality. Nonetheless, these methods generally leave the original circuit's layout visible, allowing an attacker to discern the boundary between the original circuit and the inserted random circuit. Methods introduced in \cite{topaloglu2023quantum, liu2025eloqenhancedlockingquantum} involved concepts similar to classical logic locking, utilizing numerous key bits to manage the circuit's operations. Work in \cite{rehman2025opaque} presents a methodology to encode the rotation angles of existing phase gates using key bits and to create dummy phase gates, incorporating a large key space with minimal cost.

\section{Threat Models}
\begin{figure}[htb]
\captionsetup{font=small} 
    \centering
    \includegraphics[width=0.5 \textwidth]{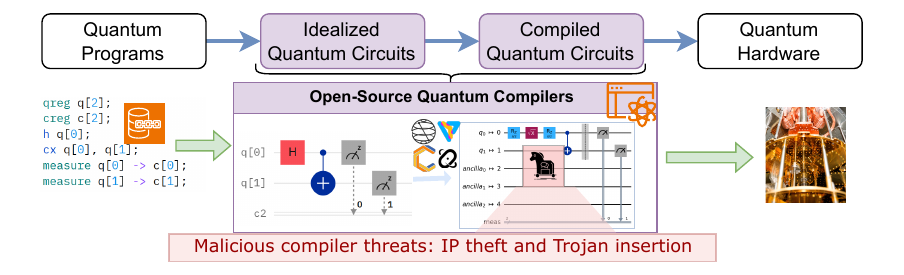}
    \caption{Illustration of the untrusted compiler threat model, including IP theft and Trojan insertion.}
    \label{fig:threat_model}
    \vspace{-3mm}
\end{figure}

In this study, we regard third-party quantum compilers as potential threats. These third-party quantum compilers provide various advantages, including support for diverse quantum computing platforms, enhanced optimization, and error correction \cite{smith2020open,salm2021automating}. Prominent third-party quantum compilers include Qulic \cite{smith2020open} and TKET \cite{sivarajah2020t}, among others. Additionally, certain hardware-specific compilers like IBM's Qiskit \cite{qiskit2024} and Google's Cirq \cite{hancockcirq} can also operate across different platforms, acting as third-party solutions. It is important to emphasize that these examples only serve to illustrate the widespread use of third-party compilers, and we do not imply any malicious intent on their part. Despite offering flexibility, third-party compilers pose challenges to the protection of intellectual property in quantum circuits. During compilation, the entire quantum circuit design is revealed, making its IP susceptible to unauthorized use or reproduction \cite{ghosh2023primer}. Unauthorized individuals gaining access could duplicate the original design and replicate its functionality. Figure \ref{fig:threat_model} presents these threat models. The threat model explored in this paper aligns with those presented in prior quantum adversarial research \cite{das2023randomized, suresh2021short, das2024secure,topaloglu2023quantum, saki2021split}.

\vspace{-4mm}

\section{Combined Logic and Angle Obfuscation}

The aim of the \acronym\ framework is to achieve high error impact with wrong keys and a large key search space simultaneously to IP exposure. The strategy is to encode logic and phase angle using key bits in the obfuscated quantum circuits. 
For logic obfuscation, we insert an ancilla qubit initialized with $\ket{0}$ and a Hadamard $H$ gate on the ancilla qubit to separate key bits.
Phase obfuscation can be achieved by using $R_Z$ phase gates to obfuscate the functionality of quantum circuits and the rotation angle is the secret key. 
Obfuscation is performed before compilation.
After the circuit has been compiled, a de-obfuscation procedure is performed to restore the correct functionality.
Each $H$ gate on the ancilla qubit is either removed or replaced with an $X$ gate to represent the correct key. Obfuscated rotation angles in the phase gates will be replaced with the correct angle.
In this section, we introduce the obfuscation and de-obfuscation procedures.

\subsection{Obfuscation Procedure} \label{ssec:obf}
Choosing the obfuscation location for both components of key-gates and determining the rotation angles for locked phase gates are the two primary tasks in the obfuscation process. In quantum circuit obfuscation, the optimal placement for obfuscation should introduce the highest level of error at the circuit's output qubits. In classical circuit logic locking, key gates are most effective when positioned at nodes whose transitive fan-out cone encompasses a large number of output bits \cite{yasin2015improving}. Analogously, for \acronym, locating areas that significantly affect output errors is crucial when inserting key gates.
To this end, we organize quantum gates into layers and make sure that each layer only has non-phase gates or phase gates (not a combination of them). 
For the non-phase and phase layers, we apply the logic and phase obfuscation procedures, respectively.
The detailed process is 
given in Algorithm \ref{alg:lock}. 


\begin{algorithm}[htb]
\DontPrintSemicolon
\SetAlgoLined
\SetNoFillComment
\LinesNotNumbered 
\caption{The \acronym\ Obfuscation Procedure}\label{alg:lock}

\KwIn{Quantum circuit $Q$}
\KwOut{Obfuscated quantum circuit $Q'$, correct key $\vec{k}$}


\tcc{Layer Splitting}
Split each layer into:
\begin{itemize}
    \item Phase gate layer (e.g., $Rz, P, S, T$)
    \item Non-phase gate layer (e.g., $X, Y, Z, H, CX$)
\end{itemize}

\tcc{Identify Obfuscation Locations}
Select obfuscation locations for logic $\mathcal{L}_L$ and phase $\mathcal{L}_P$, including existing gates and empty slots.\;

\tcc{Logic Obfuscation}
Add ancillary qubit $q_k$ for logic keys.\\
\ForEach {Location in $\mathcal{L}_L$}{
    Add a Hadamard gate on $q_k$.\\
    \If{Location contains a gate}{
        Convert the gate to a controlled gate by $q_k$.\\
        Append $\vec{k}$ with 1'b1.
    }
    \Else{
        Add a dummy gate controlled by $q_k$.
        Append $\vec{k}$ with 1'b0.
    }
}
\tcc{Phase Obfuscation}
\ForEach {Location in $\mathcal{L}_P$}{
    \If{Location contains a phase gate}{
        Calculate correct key with original phase angle $\alpha$: $\kappa = \alpha/\frac{\pi}{4}$.\\
        Randomize the rotation angle.\\
        Append $\vec{k}$ with the binary form of $\kappa$.
    }
    \Else{
        Add a dummy phase gate with a randomized rotation angle.\\
        Append $\vec{k}$ with 3'b000.
    }
}


\KwRet{Obfuscated quantum circuit $Q'$, correct key $\vec{k}$}
\end{algorithm}

\subsubsection{Logic Obfuscation}
We first add an ancilla qubit $q_k$ to host the key bits. For each key bit, we place an $H$ gate for each logic key bit so that no information is revealed about the correct key value or the correlation between adjacent key bits.
Each key bit corresponds to a section on $q_k$ and controls a gate in one of the non-phase layers. If a key bit is $\ket{1}$, the controlled gate is formed by converting an existing non-phase gate into a controlled one. If $\ket{0}$, the controlled gate should not have an effect when the circuit is de-obfuscated, so we insert dummy controlled gates instead. We usually convert $X$ to $CX$ gates or convert $CX$ to $CCX$ gates to ensure large error impacts and achieve structural obfuscation.

\subsubsection{Phase Obfuscation}

The phase rotation angle for a gate can assume any value within the interval $[0,2\pi)$. Nevertheless, certain angles have a higher probability of being the true angle. For instance, the $S$ and $T$ gates adjust the phase by amounts of $\frac{\pi}{2}$ and $\frac{\pi}{4}$, respectively. These gates are part of the Clifford group \cite{tolar2018clifford}, a widely utilized collection in quantum algorithms. By combining these gates, one can achieve phase changes in increments of $\frac{\pi}{4}$. Consequently, these particular angles are more probable as correct guesses, enhancing the success likelihood for an adversary. Within the range $[0,2\pi)$, there are 8 increments of $\frac{\pi}{4}$. Thus, each phase gate has the capacity to encode 3 Boolean key bits. It is important to recognize that this is a conservative estimate since actual phase change angles are not exclusively restricted to multiples of $\frac{\pi}{4}$.

\vspace{-1mm}
\begin{figure*}[htp]
\captionsetup{font=small} 
    \centering
    \includegraphics[width=\textwidth]{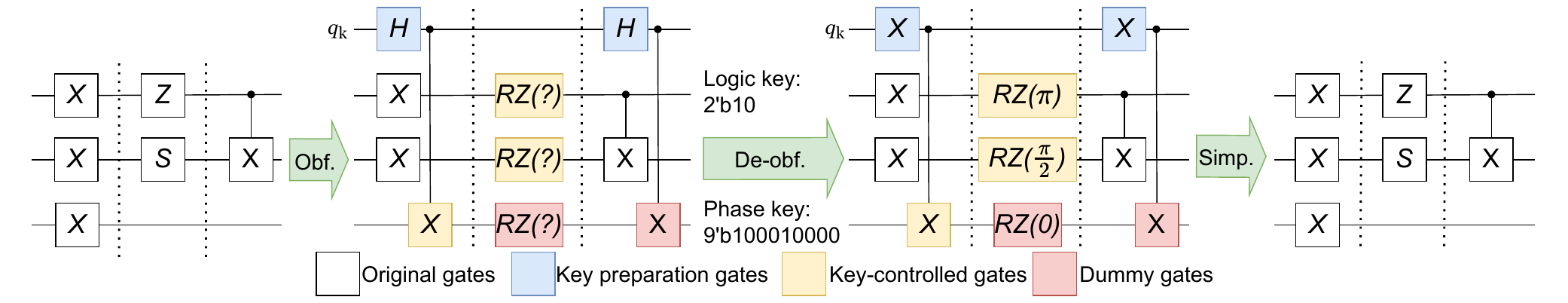}
    \vspace{-4mm}
    \caption{An Example of \acronym\ Obfuscation and De-obfuscation.}
    \label{fig:defense_flow}
\vspace{-6mm}
\end{figure*}

\subsection{De-obfuscation Process}

De-obfuscation is also performed separately for the logic and phase portions.
For the logic portion, the Hadamard gates on $q_k$ are removed and Pauli-$X$ gates are inserted at appropriate locations on $q_k$ to ensure the correct functionality after unlocking. If the key $\vec{k}$ starts with a 1, a Pauli-$X$ gate is needed to prepare $q_k$ in the state $\ket 1$. If the next key bit is 0, another Pauli-$X$ gate toggles $q_k$'s state back to $\ket 0$. In general, a Pauli-X gate is inserted on $q_k$ whenever the next key bit differs from the current one. Using this process, the unlocked circuit can be further simplified because the behavior of the controlled gates is already determined by the key. Gates controlled by $\ket 1$ can be replaced with an independent gate, and those controlled by $\ket 0$ can be removed. The key qubit $q_k$ can also be removed as it is not part of the original circuit.
For the phase obfuscation, the rotation angle of the obfuscated phase gates will be set by the key bits assigned to the gate. Specifically, the rotation angle will be set to $\frac{\kappa\pi}{4}$, where $\kappa$ stands for the integer between 0 and 7 that is encoded by the three key bits assigned to the gate. If this key matches the locking key, the correct phase angles will be restored, and the de-obfuscated compiled quantum circuit will have correct functionality when executed on quantum computers. If an incorrect de-obfuscation key is used, the resulting circuit will experience significant functional corruption, as we will describe in our experiment results.

Figure \ref{fig:defense_flow} illustrates the obfuscation and de-obfuscation flow. The dotted lines marks layer boundaries, which are marked in the original circuit when the phase and non-phase gates are split into different layers. Both existing and added dummy gates are involved in the obfuscation, which can hide both the functionality and the structure of the circuit. With the correct de-obfuscation key, the compiled circuit can be simplified to match the original circuit, eliminating all the added dummy gates and the key ancilla qubit while retaining original gates.

\section{Experiments Evaluation}
\subsection{Experimental Setup}

For the experiments, the IBM Qiskit framework was used to compile and simulate quantum circuits, and the benchmarks were taken from the \textit{Qasm} benchmark catalog \cite{li2023qasmbench}. These benchmarks have been utilized in previous work for quantum circuit compilation. The catalog spans a wide range of quantum algorithms and gate operations with a typical qubit size ranging from 4 to 96. 
In our case, smaller quantum circuits of 3 to 4 qubits were chosen to maintain a reasonable testing time and simplify data tracking. However, our method can also be applied to larger qubit circuits.

To approximate realistic simulation conditions, the \textit{FakeManilaV2} backend from Qiskit \cite{qiskit2024} is chosen, which incorporates a noise model similar to that of an actual IBM Quantum device. All simulations were performed with 1024 shots to generate statistically significant results. Moreover, both the original and the obfuscated circuits were simulated using the same backend, ensuring that any differences observed are attributable to the mechanism rather than variations in the simulation environment.

\subsection{Metrics for Evaluation}

\textit{Total Variation Distance (TVD)} serves as a metric to assess the distance between two probability distributions. This measure is particularly advantageous for evaluating quantum circuit outputs since such circuits yield probabilistic results, in contrast to the deterministic outputs of classical computers. For instance, the result of simulating a 1-bit quantum circuit with noise can be expressed as a distribution like \{``0'': 95, ``1'': 5\}, derived from 100 trials. TVD quantifies the deviation between the output distributions of two quantum circuits. When comparing the ideal circuit and an obfuscated version, TVD indicates how well the obfuscation masks the true output. It is computed by summing the absolute differences between counts for each possible outcome in the two distributions, then normalizing by the total number of trials. The TVD formula is:
\begin{equation}
    TVD = \frac{\sum_{i=0}^{2^b-1} |y_{i,a} - y_{i,b}|}{2N}
\end{equation}
Where $N$ represents the total number of shots in this run, $b$ represents the number of output qubits, resulting in $2^b$ possible output types. $y_{i,a}$ and $y_{i,b}$ represent the count of value $i$ in the two quantum circuits being compared. 

Previous research on quantum circuit obfuscation used the quantum circuits directly for TVD analysis, which essentially means that they only considered the all-$\ket{0}$ input. 
In reality, quantum circuits may have any input state, just like classical circuits can have different input patterns.
In order to evaluate the overall functional impact of the obfuscation approach, we sample the quantum circuit input space by adding a $U_3$ gate at the input side of each qubit.
Through implementing independent $U_3$ rotations onto each qubit, the reachable state space is increased. After the initial $U_3$ layer, we apply the obfuscation of logic and phase angles to obfuscate the circuit. 
In our experiments, for each circuit, we randomly sample 10 different input states and obtain the output state distribution with 100 shots.
The de-obfuscation reverses the changes made in the obfuscation using the correct key, returning the output to the appropriate state. 


\begin{table}[tb]
\centering
\caption{Circuit depth and gate count changes during obfuscation and the numbers of key bits incorporated.}
\setlength{\tabcolsep}{0.5pt} 
\begin{tabular}{|l|p{7mm}|p{8mm}|p{10mm}|p{10mm}|p{13mm}|p{14mm}|}
\hline

\textit{Circuit} & 
\textit{Depth} & 
\textit{\CellWithForceBreak{Depth \\ Obf.}}& 
\textit{\CellWithForceBreak{Org. \\ \# Gates}}  & 
\textit{\CellWithForceBreak{Obf. \\ \# Gates}} & 
\textit{\CellWithForceBreak{Logic Key \\ \# Bits}} & 
\textit{\CellWithForceBreak{Phase Key\\ \# Bits}} \\ \hline

Adder & 12  & 17  & 23 & 113 & 90 & 60 \\ \hline
Basis Change & 22 & 27 & 46 & 137 & 137 & 45 \\ \hline
Fredkin & 12 & 19 & 19 & 91 & 34 & 84 \\ \hline
Wstate & 6 & 11 & 6 & 44 & 38 & 45 \\ \hline

\end{tabular}

\label{tab:circuit_parameters}
\end{table}

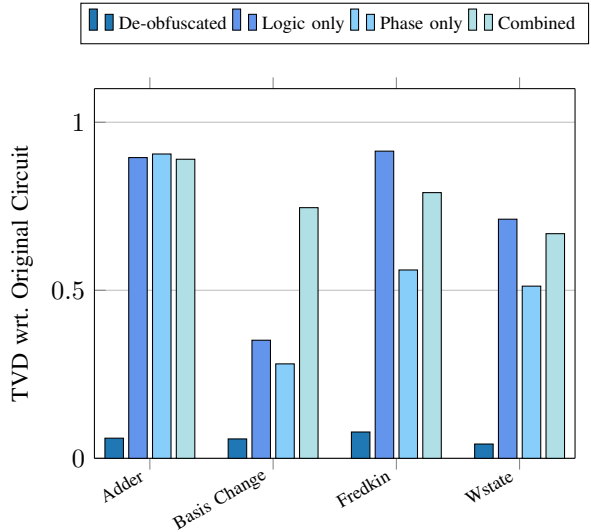
\begin{figure}[!t]
\centering
\begin{tikzpicture}
\begin{axis}[
    ybar,
    bar width=7pt,
    width=0.9\columnwidth,
    height=6.5cm,
    ylabel={\small TVD wrt. Original Circuit},
    symbolic x coords={Adder, Basis Change, Fredkin, Wstate},
    xtick=data,
    xticklabel style={rotate=30, anchor=east, font=\scriptsize},
    ymin=0, ymax=1.1,
    enlarge x limits=0.15,
    legend style={at={(0.5,1.12)}, anchor=south, legend columns=-1, font=\scriptsize},
    axis y line*=left,
    ymajorgrids,
]

\definecolor{bar1}{RGB}{31,119,180}
\definecolor{bar2}{RGB}{100,149,237}
\definecolor{bar3}{RGB}{135,206,250}
\definecolor{bar4}{RGB}{176,224,230}

\addplot[fill=bar1] coordinates {(Adder,0.0599) (Basis Change,0.0577) (Fredkin,0.0783) (Wstate,0.0425)};
\addplot[fill=bar2] coordinates {(Adder,0.8948) (Basis Change,0.3514) (Fredkin,0.9140) (Wstate,0.7115)};
\addplot[fill=bar3] coordinates {(Adder,0.9055) (Basis Change,0.2810) (Fredkin,0.5604) (Wstate,0.5123)};
\addplot[fill=bar4] coordinates {(Adder,0.8899) (Basis Change,0.7458) (Fredkin,0.7906) (Wstate,0.6684)};

\legend{De-obfuscated, Logic only, Phase only, Combined}
\draw[line width=0.4pt] (rel axis cs:1,0) -- (rel axis cs:1,1);

\end{axis}
\end{tikzpicture}
\caption{TVD Comparison across circuits.}
\label{fig:tvd_dualaxis_ieee}
\end{figure}

\subsection{Result Analysis}
In this section, we present our experimental results, using the \textit{Qasm} benchmarks simulated using the Qiskit backend. The benchmarks that were selected for this experiment include \textit{Adder}, 
\textit{Fredkin}, \textit{Basis Change}, and \textit{Wstate}. These circuits have a range of complexities and structures that allow us to evaluate performance across a variety of scenarios.
The impact of obfuscation on circuit behavior is captured in the TVD (Total Variation Distance) analysis shown in Figure \ref{fig:tvd_dualaxis_ieee}. We use the original circuit as the golden reference and measure its TVD with the obfuscated (logic, phase, combined) and the restored circuits. As expected, locked circuits exhibit a high TVD, reflecting the disruptive effect of incorrect keys. After de-obfuscation, TVD values drop significantly, demonstrating that correct-key restoration successfully recovers expected circuit behavior.
Among the tested circuits, \textit{Adder} and \textit{Fredkin} show the highest locked-state TVD values (0.8943 and 0.9125, respectively), consistent with the presence of multiple phase gates and conditional logic structures.

\subsection{Cost and Overhead Analysis}

Table~\ref{tab:circuit_parameters} summarizes the number of key bits incorporated into the obfuscated circuits and the changes in circuit parameters after applying both logic and phase obfuscation. Notice that the depth and gate count increase are \textit{NOT} obfuscation overhead because they are temporary. As depicted in Figure \ref{fig:defense_flow}, when the circuit is de-obfuscated with the correct key, the circuit can be simplified and the added key ancilla qubit and the dummy gates are removed. Nonetheless, a large key space can be incorporated using the proposed obfuscation framework, which ensures the original circuit is sufficiently obfuscated during compilation.
Compared to existing approaches using only one type of obfuscation, the number of key bits incorporated through the other type of obfuscation represents the security improvement in our approach over the existing ones.
The obfuscated circuits displayed a high TVD relative to the original output, which confirms that a high level of obfuscation was achieved through logic and phased locking. After applying the de-obfuscation procedure, the restored circuits recovered a high fidelity, with only about 5\% TVD w.r.t. the original circuit on average. The primary causes of this difference are that, when the original circuit is split layers, we need to insert a barrier into the circuit, which the compiler is prevented from optimizing the circuit across. This leads to some missed opportunities in the optimizations applied to the compiled circuits, which has a noticeable effect on the circuit accuracy in the current noisy quantum hardware.

These results show that the obfuscation technique causes major functional distortion with incorrect keys but nearly restores the original behavior after de-obfuscation.

\section{Conclusion}
We put forward \acronym, a new approach to the quantum circuit obfuscation framework to examine how combining logic and phase key control could enhance circuit obfuscation without significantly affecting recoverability. Our approach separates circuits into phase and non-phase layers, applies XOR logic locking and key-dependent phase rotations, and incorporates randomized single qubit gates to increase uncertainty before compilation. We also introduced an input state sampling evaluation method to better capture the overall functional effects of obfuscation. While the current results demonstrate that this combined approach can induce measurable disruption under incorrect keys and maintain accuracy under correct de-obfuscation, this study primarily serves as an initial exploration.

\section*{Acknowledgment}
This work is supported by the National Science Foundation under Award 2530705.

\balance

\newpage

\bibliographystyle{IEEEtran.bst}
\bibliography{qref}
\end{document}